\documentclass[conference]{IEEEtran}
\IEEEoverridecommandlockouts
\usepackage{cite}
\usepackage{amsmath,amssymb,amsfonts}
\usepackage{algorithmic}
\usepackage[linesnumbered,ruled,noend]{algorithm2e}
\usepackage{graphicx}
\usepackage{textcomp}
\usepackage{xcolor}
\usepackage{caption}
\usepackage{subcaption}
\usepackage{array}
\newcolumntype{L}[1]{>{\raggedright\let\newline\\\arraybackslash\hspace{0pt}}m{#1}}
\newcolumntype{C}[1]{>{\centering\let\newline\\\arraybackslash\hspace{0pt}}m{#1}}
\newcolumntype{R}[1]{>{\raggedleft\let\newline\\\arraybackslash\hspace{0pt}}m{#1}}

\usepackage{multicol}
\usepackage{multirow}
\usepackage{enumitem}
\usepackage{pdfpages}
\usepackage{booktabs}
\usepackage{url}
\def\checked{\checkmark}
\def\notchecked{$\times$}
\def\notapplicable{n/a}
\def\HiLiYellow{\leavevmode\rlap{\hbox to \hsize{\color{yellow!50}\leaders\hrule height .8\baselineskip depth .5ex\hfill}}}
\def\HiLiRed{\leavevmode\rlap{\hbox to \hsize{\color{red!30}\leaders\hrule height .8\baselineskip depth .5ex\hfill}}}

\def\usericon{\includegraphics[height=\fontcharht\font`\B]{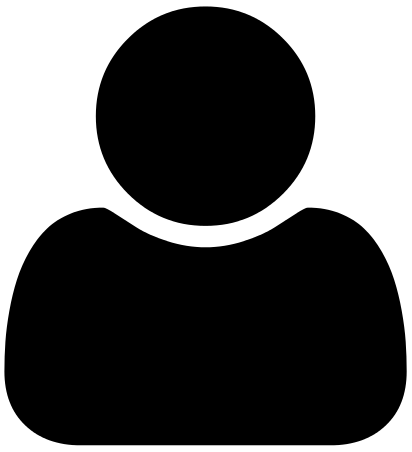}}
\def\BibTeX{{\rm B\kern-.05em{\sc i\kern-.025em b}\kern-.08em
    T\kern-.1667em\lower.7ex\hbox{E}\kern-.125emX}}

\newcommand\blfootnote[1]{%
  \begingroup
  \renewcommand\thefootnote{}\footnote{#1}%
  \addtocounter{footnote}{-1}%
  \endgroup
}
\newtheorem{definition}{Definition}
    
\begin{document}

\title{Efficient Strategies for Graph Pattern Mining Algorithms on GPUs}

\author{
    \IEEEauthorblockN{
        Samuel Ferraz\IEEEauthorrefmark{1}\IEEEauthorrefmark{2},
        Vinicius Dias\IEEEauthorrefmark{1}\IEEEauthorrefmark{3},
        Carlos H. C. Teixeira\IEEEauthorrefmark{1}, George Teodoro\IEEEauthorrefmark{1}, 
        Wagner Meira Jr.\IEEEauthorrefmark{1}
    }
    \IEEEauthorblockA{\IEEEauthorrefmark{1} \textit{Computer Science Department, Federal University of Minas Gerais (UFMG) — Belo Horizonte, Brazil} \\
    \textit{Email: \{samuel.ferraz, viniciusvdias, carlos, george, meira\}@dcc.ufmg.br}
    }
    \IEEEauthorblockA{\IEEEauthorrefmark{2} \textit{School of Computing, Federal University of Mato Grosso do Sul (UFMS) — Campo Grande, Brazil} \\ 
    \textit{Email: samuel.ferraz@ufms.br}
    }
    \IEEEauthorblockA{\IEEEauthorrefmark{3} \textit{Department of Computing and Systems, Federal University of Ouro Preto (UFOP) — João Monlevade, Brazil} \\ 
    \textit{Email: viniciusvdias@ufop.edu.br}
    }
}

\maketitle

\begin{abstract}
Graph Pattern Mining (GPM) is an important, rapidly evolving, 
and computation demanding area. GPM computation relies on subgraph enumeration, which consists in extracting subgraphs that match a given property from an input graph. Graphics Processing Units (GPUs) have been an effective platform to accelerate applications in many areas. However, the irregularity of subgraph enumeration makes it challenging for efficient execution on GPU due to typical uncoalesced memory access, divergence, and load imbalance. Unfortunately, these aspects have not been fully addressed in previous work. Thus, this work proposes novel strategies to design and implement subgraph enumeration efficiently on GPU. We support a depth-first search style search (DFS-wide) that maximizes memory performance while providing enough parallelism to be exploited by the GPU, along with a warp-centric design that minimizes execution divergence and improves
utilization of the computing capabilities. We also propose a low-cost load balancing layer to avoid idleness and redistribute work among 
thread warps in a GPU. Our strategies have been deployed in a system named DuMato, which provides a simple programming interface to allow efficient implementation of GPM algorithms. Our evaluation has shown that DuMato is often an order of magnitude faster than state-of-the-art GPM systems and can mine larger subgraphs (up to 12 vertices).

\end{abstract}


\section{Introduction} \label{section:introduction}

Graph pattern mining (GPM)\blfootnote{$\copyright$ 2022 IEEE. Personal use of this material is permitted.  Permission from IEEE must be obtained for all other uses, in any current or future media, including reprinting/republishing this material for advertising or promotional purposes, creating new collective works, for resale or redistribution to servers or lists, or reuse of any copyrighted component of this work in other works.} aims to unveal relevant subgraph patterns in graphs, being widely used in different domains and applications from social media~\cite{DT14} to biological networks analysis~\cite{SK04}. It relies on subgraph enumeration over an input graph, which consists of visiting subgraphs that match a desired graph property. Subgraph enumeration incurs in high computational 
cost and memory demands as the size of the mined subgraphs increases~\cite{LXXL15,RSL12,DTGMP19,JMV20}. For example, the small biological dataset \textit{bio-diseasome}
\footnote{
\url{https://networkrepository.com}
}
(516 vertices, 1.2K edges) contains 112B induced subgraphs with 10 vertices, which would require around 4~TB of memory with a 4-byte integer per vertex to store all subgraphs.

\begin{figure}[!hbt]
\centering
\includegraphics[width=.9\linewidth]{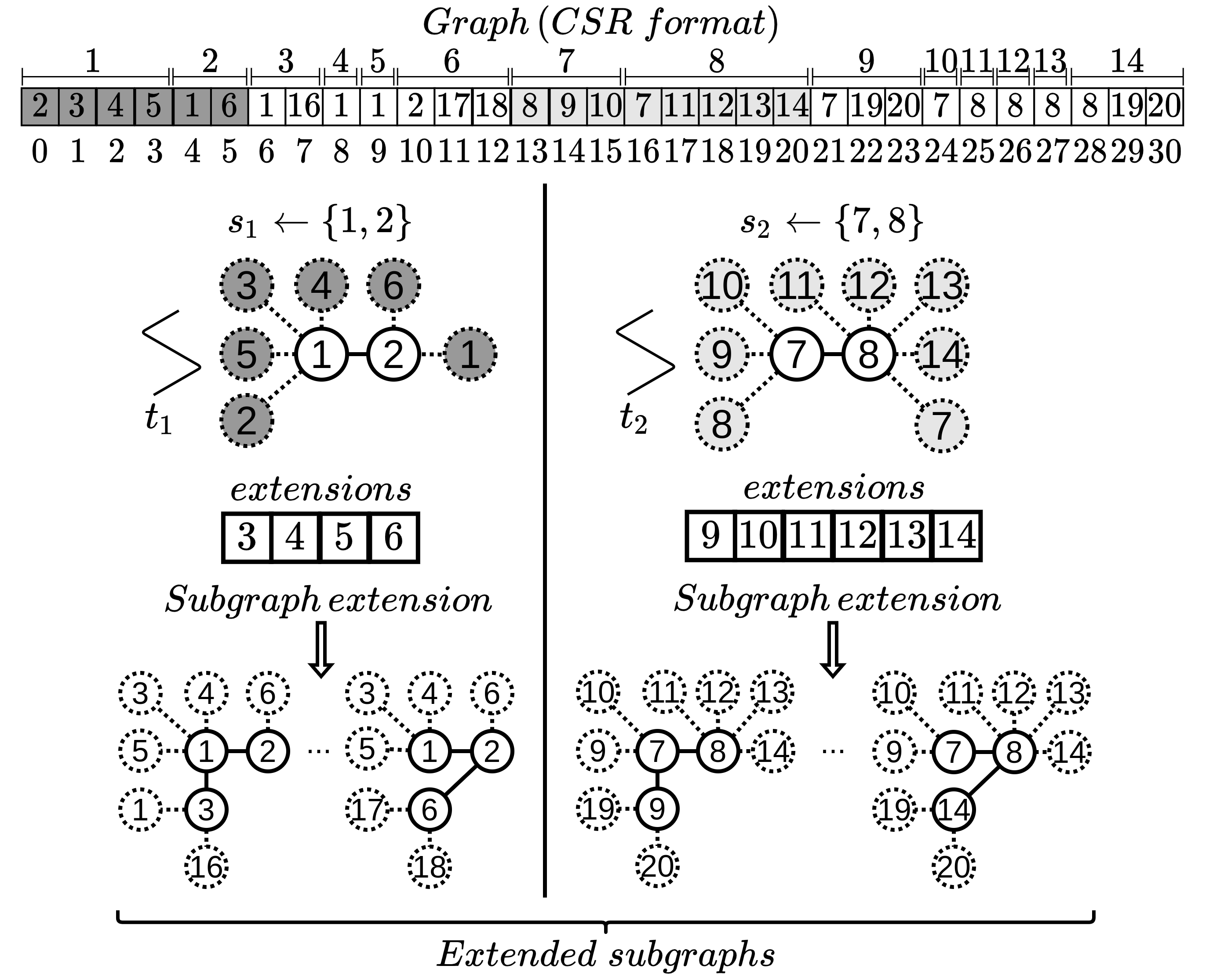}
\caption{Subgraph-centric parallel processing of subgraphs.
}
\label{figure:challenges_example}
\end{figure}

A core operation in subgraph enumeration is the \emph{subgraph extension}, in which subgraphs with $k+1$ vertices are obtained by the combination of a subgraph $s$ with $k$ vertices with a set of extensions (vertex ids) derived from the adjacency of vertices in $s$.  Figure~\ref{figure:challenges_example} illustrates this process using induced subgraphs $s_1$ and $s_2$. Subgraph extension of $s_1$ generates four extended subgraphs while $s_2$, six. The huge amount of subgraphs to be explored may cause long execution times, leading the pursuit of massively parallel architectures.

Parallel implementations of specific GPM algorithms were proposed for shared and distributed memory machines~\cite{guo2020subgraphmatchinggpu,lai2016seed,ribeiro2014gtries,abdelhamid2016scalemine,danisch2018kclist,eppstein2013maximalcliques}.
These solutions usually relax the definition of subgraph enumeration for each algorithm, reducing the computational cost and memory demands.
Although they have attained good performance on some settings, the
implementation of new algorithms was still a challenging and laborious process, as the restrictions used to design one GPM algorithm may not apply to others. This motivated the development of GPM systems, which support subgraph enumeration and allow the implementation of specific GPM algorithms by supporting custom graph properties. GPM systems offer a good tradeoff between
programmability and performance~\cite{DTGMP19,JMV20,TFSSZA15,MW19,WZTNX18}.

Graphics Processing Units (GPUs) are successful in accelerating applications in many domains. However, the parallelization strategies for subgraph enumeration on GPU present major limitations concerning efficient use of GPU architecture, especially with regard to \textit{memory uncoalescence}, \textit{divergences} and \textit{load imbalance}. Pangolin~\cite{CDGP20} is the only GPM system to employ GPUs. However, it has a high memory demand and has not been designed and implemented to fully utilize the GPU computing power by avoiding divergence, improving memory access pattern and mitigating load imbalance. Next we detail the main challenges concerning parallel subgraph enumeration on GPU addressed in this work.

The first challenge is the \emph{high memory demand} imposed by enumeration. As subgraph extension relies on combining a subgraph with its extensions, it may lead to combinatorial explosion in the number of subgraphs as the enumeration progresses.

A breadth-first search
(BFS) style, which is used in Pangolin~\cite{CDGP20}, is a natural choice for parallel subgraph enumeration as it
exports a regular parallelism in the exploration of adjacency lists. However, BFS materializes all the states related to extended subgraphs, and the amount of memory required by this strategy quickly grows
with the size of the subgraph, limiting its usage to enumerate only small subgraphs. 
On the other hand, depth-first search (DFS) style approach
reduces the memory demand as only a small portion of the states (subgraphs being processed) are kept during the enumeration, but its parallel performance may be severely affected on GPU by its irregular and strided memory requests.

The second challenge arises from the inherent \emph{irregularity} of enumeration. Current GPM systems implement parallel subgraph enumeration using a subgraph-centric processing, where subgraphs are treated as independent tasks~\cite{SKWNRRP14}. Consequently, the thread-based parallel exploration of Figure~\ref{figure:challenges_example} (used in Pangolin~\cite{CDGP20}), in which each thread independently explores distinct subgraphs, results in a reduced GPU performance due to the intrinsic \textit{memory uncoalescence}, thread \textit{divergences} and
\textit{load imbalance}.
\textit{Memory uncoalescence} and subutilization of the memory bandwidth arises as different threads within a warp access strided memory locations/graph portions according to their currently processed subgraph. 
For example, thread $t_1$ accesses positions [0-5] (dark gray positions) of graph when performing subgraph extension, 
while thread $t_2$ accesses positions [13-20] (light gray positions). 
\textit{Divergence} in the execution occurs due to different sizes and processing costs of the adjacency lists (very common in scale-free graphs), leading to inefficient utilization of the GPU. 

\textit{Load imbalance} is the third challenge which 
exists as the cost of exploring subgraphs may be different and can not be known in
advance. Thus, even if subgraphs $s_1$ and $s_2$ are assigned to different threads to take advantage of parallelism, idleness is likely to occur as some threads finish earlier. In this paper we propose strategies to mitigate these three challenges and allow efficient implementation of GPM algorithms on GPUs. Our contributions are summarized as:

\begin{itemize}[wide, labelwidth=!, labelindent=0pt]
    \item \textbf{DFS-wide Subgraph Exploration.} Subgraph exploration strategy designed for GPU that reduces memory demands vs. BFS, and allows massive parallel exploration of subgraphs with regular memory accesses. Our DFS-wide strategy achieved an average speedup of 12$\times$ vs DFS. To the best of our knowledge, we are the first work to use a DFS-style subgraph exploration on GPU for GPM processing;
    
    \item \textbf{Warp-centric Design.} An efficient warp-centric design of all compute demanding stages of the subgraph enumeration, improving memory coalescence and minimizing divergences; 
    
    \item \textbf{Warp-Level Load-Balancing.} A load redistribution strategy among GPU warps. It is executed by CPU that monitors GPU occupancy and decides when and how load is redistributed. This layer leads to an average speedup of 16$\times$;
    
    \item \textbf{DuMato GPU-based GPM System.}  A runtime system named \emph{DuMato}, which offers a high-level API to develop GPU GPM algorithms used to deploy our optimizations. DuMato is able to explore larger subgraphs (up to 12 vertices) and is often an order of magnitude faster than state-of-art GPM systems.
\end{itemize}

\section{Background} \label{section:background}

We assume for sake of simplicity undirected graphs without labels, but
our strategies may be adapted to support directed graphs and labeled features. The vertices and edges of a graph $G$ are denoted by $V(G)$ and $E(G)$, respectively. The core of our problem lies on the enumeration of \textit{induced subgraphs}, that is, 
a subgraph $S$ where, for any  $v_i,v_j \in V(S)$, $(v_i,v_j)\in E(S)$ iff $(v_i,v_j) \in E(G)$.

A graph is explored through incremental visits to vertices' neighbourhood (Definition~\ref{definition:neighbourhood}), called \textit{traversals} (Definition~\ref{definition:traversal}). A traversal can be used to create an induced subgraph from its vertices, called \textit{induced traversal} (Definition~\ref{definition:traversal}). GPM algorithms usually traverse the graph starting from each vertex/edge, and traversal strategies are usually categorized as either \emph{breadth-first search} (\textit{BFS}) or \emph{depth-first search} (\textit{DFS}). When two traversals find an \emph{isomorphism} (Definition~\ref{definition:isomorphism}) between subgraphs sharing the same vertex set, there is an \textit{automorphism} (Definition~\ref{definition:isomorphism}).

\begin{definition}\label{definition:neighbourhood}
Given a graph $G$ and a subgraph $S$ of $G$, the \textbf{neighbourhood} of $S$ is defined as
$N(S)=\{v \in neighbours(u) \mid u \in V(S)\} \setminus V(S)$.

\end{definition}

\begin{definition}\label{definition:traversal}
    A \textbf{traversal} is an array $tr = [v_1, \cdots, v_k]$  of $k$ unique vertices of a graph $G$ ($tr \subseteq V(G)$), which stores an order each vertex $v \in tr$ is visited in $G$. An \textbf{induced traversal} is accompanied by the existing edges among the vertices.
\end{definition}

\begin{definition}\label{definition:isomorphism}
An \textbf{isomorphism} between two graphs $G$ and $H$ is a bijective function $f: V(G) \rightarrow V(H)$ such that, for all edges $(v_i,v_j) \in E(G)$, $(f(v_i),f(v_j)) \in E(H)$. An isomorphism between two graphs $G$ and $H$ such that $V(G) = V(H)$ is an \textbf{automorphism}.
\end{definition}

Given a graph $G$ and the set of traversals $T$ such that each $tr \in T$ creates the same induced traversal $tr_{ind}$, the \textit{canonical candidate} is the only traversal $tr \in T$ whose visiting order of vertices is allowed to reach $tr_{ind}$ in $G$.
GPM algorithms generate only canonical candidates, preventing different traversals from finding the same induced traversal during exploration, avoiding redundant computation. Canonical candidates may be converted to a unique representation called \textit{canonical representative} (also referred in this work as \emph{patterns}). The conversion of an induced traversal to its corresponding canonical representative is known as \emph{canonical relabeling}, which is an operation performed very often in GPM algorithms to categorize subgraphs.

GPM algorithms rely on enumerating $k$-vertex subgraphs that follow a given property. Equation~\ref{eq:enumeration-function} describes \textit{enumeration function E}, which can be used to design GPM algorithms. Given a graph $G$, an initial traversal $tr$, and an integer $k$, the function $E$ explores traversals with $k$ vertices that satisfy a given anti-monotonic property $P$, producing results through an output function $A$. Functions $P$ and $A$ enable application-specific semantics.
For instance, motif counting \cite{ribeiro2014gtries} would have $A$ counting how many canonical candidates exist per canonical representative and clique counting~\cite{danisch2018kclist} would have $P$ selecting only canonical candidates that are fully
connected.

{\small
\begin{equation}
  E(G,tr,k) =
  \left\{\begin{array}{l @{\quad} l r l}
    \emptyset & \text{if}\; |tr| = 0 
    \\
    \bigcup\limits_{u\in N(tr)}E(G,P(tr+u),k) & \text{if}\; |tr| < k 
    \\
    A(tr) & \text{if}\; |tr| = k 
    \\
  \end{array}\right.
  \label{eq:enumeration-function}
\end{equation}
}

\section{Related Work}
\label{section:related_work}

This section presents the closest GPM systems available in the literature. Table~\ref{tab:related-work-analysis} summarizes their main features.
\begin{table}[!htb]
    \centering
    \footnotesize
    \resizebox{\columnwidth}{!}{
    \begin{tabular}{l c c c c c }
        \toprule
        \multirow{2}{*}{GPM System} & Algorithmic  & \multirow{2}{*}{Proc.}  & Explor.     & Warp- & Load \\
         & approach     &       &  strate.  & centric & bal.

        \\\hline
        Arabesque~\cite{TFSSZA15} & Pattern-oblivious & CPU & BFS & \notapplicable & \checked
        \\
        RStream~\cite{WZTNX18} & Relational & CPU & BFS & \notapplicable & \checked
        \\
        Automine~\cite{MW19} & Pattern-aware & CPU & DFS & \notapplicable & \notchecked
        \\
        Fractal~\cite{DTGMP19} & Pattern-oblivious & CPU & DFS & \notapplicable & \checked
        \\
        Peregrine~\cite{JMV20} & Pattern-aware & CPU & DFS & \notapplicable & \notchecked
        \\
        Pangolin~\cite{CDGP20} & Pattern-oblivious & GPU & BFS & \notchecked & \notchecked
        \\
        DuMato & Pattern-oblivious & GPU & DFS & \checked & \checked
        \\
        \bottomrule
    \end{tabular}}
    \caption{Related work. ``n/a'' stands for ``not applicable''.
    }
    \label{tab:related-work-analysis}
\end{table}

\textbf{GPM systems for CPU.}
\textit{Arabesque}~\cite{TFSSZA15} is one of the first  GPM systems targeting distributed memory machines. The algorithmic approach adopted by Arabesque is known as pattern-oblivious because it does not rely on pattern generation to guide the subgraph enumeration. Arabesque proposes a data structure to compress subgraphs in-memory and to mitigate the memory demands of the BFS-style exploration while it also employs load balancing. 
\textit{RStream}~\cite{WZTNX18} is a relational GPM system that relies on expensive join operations to perform subgraph enumeration. It presents limitations caused by high memory consumption as the length of enumerated subgraphs increases.
\textit{Fractal}~\cite{DTGMP19} is a distributed memory CPU-based GPM system that uses a DFS exploration strategy to reduce memory demands. Fractal proposes and implements a hierarchical work-stealing mechanism to mitigate load imbalance. 
\textit{AutoMine}~\cite{MW19} proposes an automated code generation for GPM algorithms on CPU. It employs efficient scheduling of intersect/subtract operations to automate code generation for custom patterns. Because this approach is specialized for specific patterns, it may be too expensive in general-purpose GPM scenarios, where subgraph exploration typically involves multiple patterns.
\textit{Peregrine}~\cite{JMV20} is a parallel GPM system designed for shared-memory CPU machines. Both Peregrine and AutoMine use an exploration strategy known as pattern-aware, where canonical representatives are used to guide the subgraph enumeration by leveraging specialized execution plans. Although pattern-aware exploration is efficient for enumerating small subgraphs, it has limitations whenever the application searches for a large number of canonical representatives (e.g., counting large motifs). DuMato adopts a pattern-oblivious exploration strategy and, thus, is not limited by the number of patterns mined.

\textbf{GPM systems for GPU.} \textit{Pangolin}~\cite{CDGP20} is the only GPM system designed for GPU and  follows a pattern-oblivious enumeration using the BFS exploration strategy. Pangolin's design enables execution optimizations by pruning the search-space of subgraphs and by reducing the amount of isomorphism tests required. Materialized intermediate states generated by the BFS exploration facilitate the runtime to leverage BSP/CGM load balancing schemes. However, the BFS high memory demand limits its applicability to enumerate small subgraphs. Besides, Pangolin does not leverage optimizations to handle irregularity of parallel GPM algorithms on GPU and relies on CPU frameworks to perform isomorphism tests. 

As presented in Table~\ref{tab:related-work-analysis}, DuMato is the only GPM system designed to efficiently use GPU with a DFS-like exploration strategy. Consequently, the memory demands, compared to Pangolin, are significantly smaller and larger subgraphs can be mined. Our system also computes isomorphism tests efficiently on GPU.
DuMato has been optimized to use the warp-centric execution model, leading to better use of the GPU computing power due to the reduced thread warp divergence and optimized (coalesced) memory accesses. Finally, DuMato also proposes and implements a lightweight load balancing mechanism to redistribute load among GPU thread warps.

\section{Efficient Strategies for GPU Graph Pattern Mining}

In this section we present our strategies for efficient graph pattern mining algorithms on GPU. We start with an overview of \textit{DuMato}, our system that supports high-level implementation of GPM algorithms on GPU. Next, we use DuMato execution workflow to present our strategies to reduce memory demand, improve memory coalescence and divergences, and mitigate load imbalance.

\subsection{DuMato Execution Workflow} \label{subsection:dumato_execution_workflow}

\begin{figure}[!htb]
\centering
\includegraphics[width=.9\linewidth]{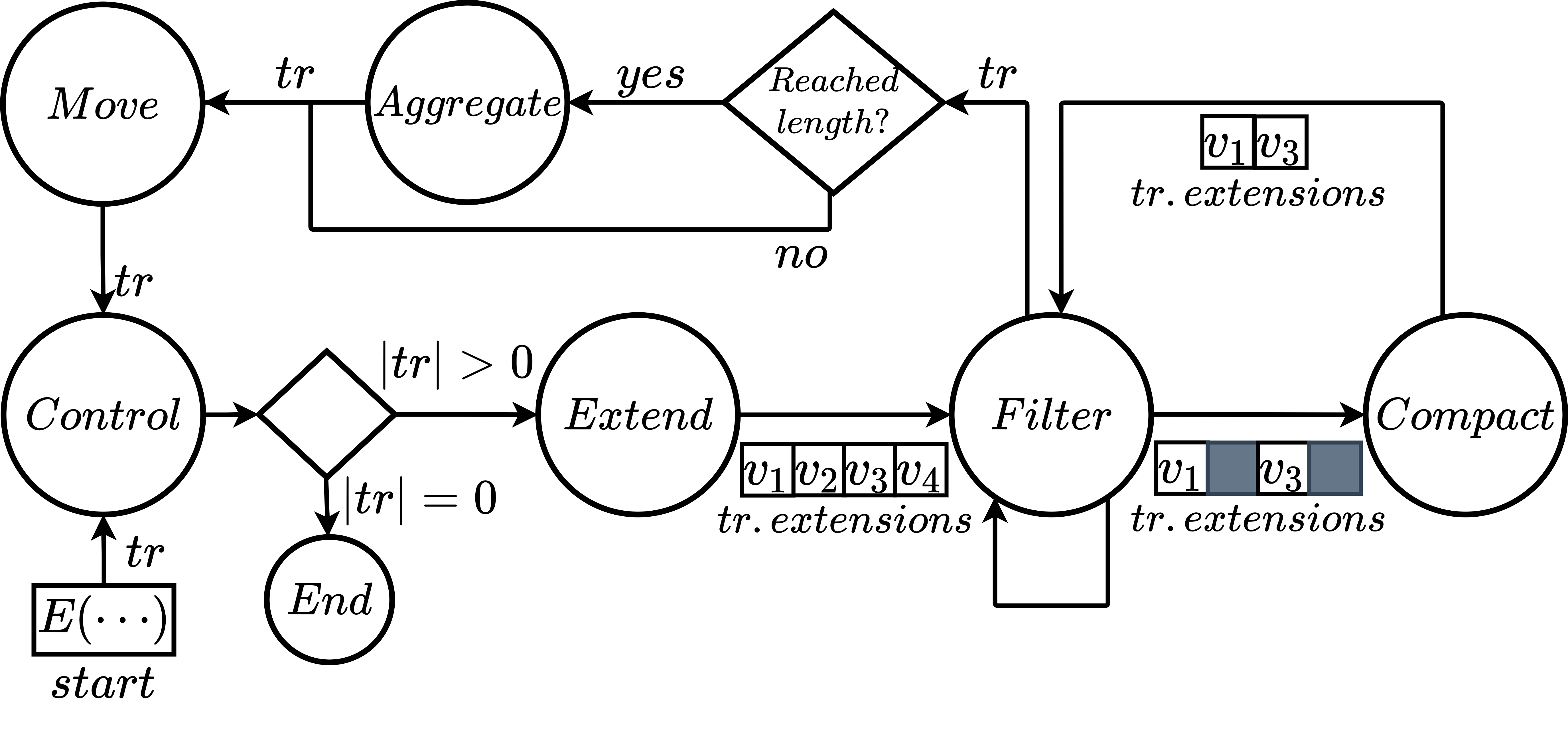}
\caption{DuMato execution workflow.}
\label{figure:execution_workflow}
\end{figure}

The execution workflow of DuMato (Figure~\ref{figure:execution_workflow}) employs the \textit{filter-process} model~\cite{TFSSZA15}, which allows 
implementation of GPM algorithms based on the enumeration function $E$
(Eq.~\ref{eq:enumeration-function}). Each circle refers to a phase 
in the workflow and each diamond refers to a decision.

The process starts with a call $E(G,tr,k,P)$ to enumerate and
output traversals of size $k$ that satisfy property $P$ extended from an initial traversal 
$tr$. The initial traversal $tr$ is 
used as input to a \emph{Control} phase, which implements the termination condition  ($|tr|=0$ in Eq.~\ref{eq:enumeration-function}). The output 
of the \emph{Control} phase is a decision on whether the subgraph enumeration should proceed (traversal is not empty) or terminate (traversal becomes empty). If enumeration continues, the \emph{Extend} phase computes
the extensions from the current traversal ($|tr| < k$ in Eq.~\ref{eq:enumeration-function}). These possible extensions are in the neighborhood of the current subgraph and are obtained 
from the adjacency of vertices in the traversal
(Def.~\ref{definition:neighbourhood}). The \emph{Extend} phase outputs the
current traversal and its extensions.

Next, application-specific semantics may be employed to narrow the subgraph search in
the \emph{Filter} phase, which selects subgraphs that satisfy some
anti-monotonic property $P$ ($|tr| < k$ in Eq.~\ref{eq:enumeration-function}). For example, a property $P$ may check whether a subgraph is a clique. This is carried out by passing over the extensions to invalidate those that do not satisfy property $P$. Multiple Filters may be executed depending on which conditions must be verified to ensure property $P$. The Filter phase outputs the current traversal and extensions that are valid.

The output of the Filter may have several invalidated (erased) positions in the array of extensions, e.g. $v_2$ and $v_4$ in Figure~\ref{figure:execution_workflow}. This array of non-contiguous valid extensions may cause substantial performance degradation, as the next Filter may have to pass over and process (or check) invalid values. Thus, DuMato proposes an optional \emph{Compact} phase executed after
each Filter to reorganize valid extensions into a contiguous
memory/array.

After all Filter/Compact phases were executed, if traversal size reaches the target number of vertices, they are forwarded to the enumeration output $A$ ($|tr| = k$ in Eq.~\ref{eq:enumeration-function}). This is accomplished in the \emph{Aggregate} phase, in which traversals are consumed for counting, pattern counting, or buffering. If the traversal has not reached the target number of vertices, the Aggregate phase is skipped.

Further, the \emph{Move} phase decides whether to move forward or backward in the subgraph enumeration. Moving forward means that an unprocessed extension is appended to the current traversal for processing (recursion call). Moving backwards means that all extensions of the current traversal have been processed, and that the algorithm can go back on processing smaller traversals (recursion return). The output of the \emph{Move} is a modified traversal that should restart the workflow at the Control, closing the cycle in Figure~\ref{figure:execution_workflow}.

Our task allocation modeling is warp-centric and each warp receives a traversal to enumerate. Threads within a warp enumerate the same traversal cooperatively, alternating between SIMD and SISD phases throughout the execution workflow. The next sections detail the design and the implementation of each phase of the execution workflow, which shows our strategies to mitigate the memory demand (Section~\ref{subsection:dfs_wide_subgraph_exploration}), the execution irregularity (Section~\ref{section:efficient_warp_centric_filter_process}), and the load imbalance (Section~\ref{subsection:out_of_gpu_load_balancing}) of GPM algorithms on GPU. We also present our easily-programmable API (Section~\ref{section:programming_dumato}).

\subsection{DFS-wide Subgraph Exploration} \label{subsection:dfs_wide_subgraph_exploration}
 
\begin{figure}[!hbt]
\begin{subfigure}{.5\textwidth}
  \centering
  \includegraphics[width=.6\linewidth]{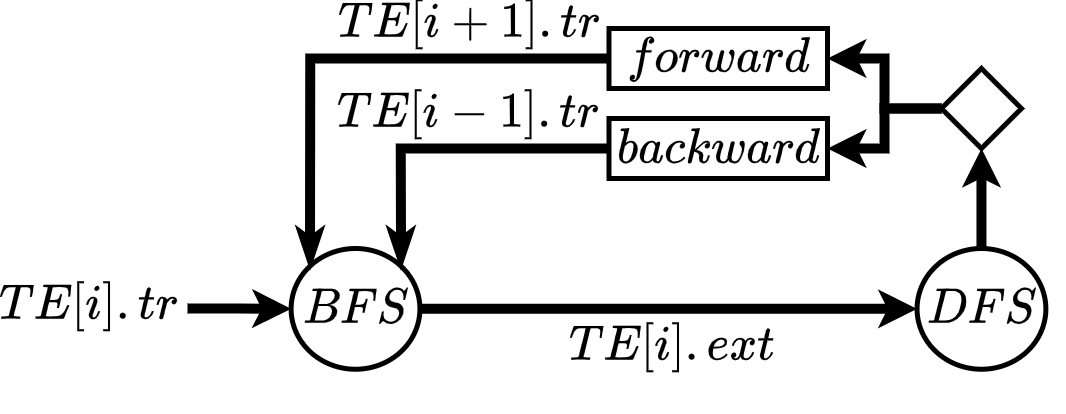}
  \caption{Overview.}
  \label{figure:DFS_wide_overview}
\end{subfigure}
\begin{subfigure}{.5\textwidth}
  \includegraphics[width=.98\linewidth]{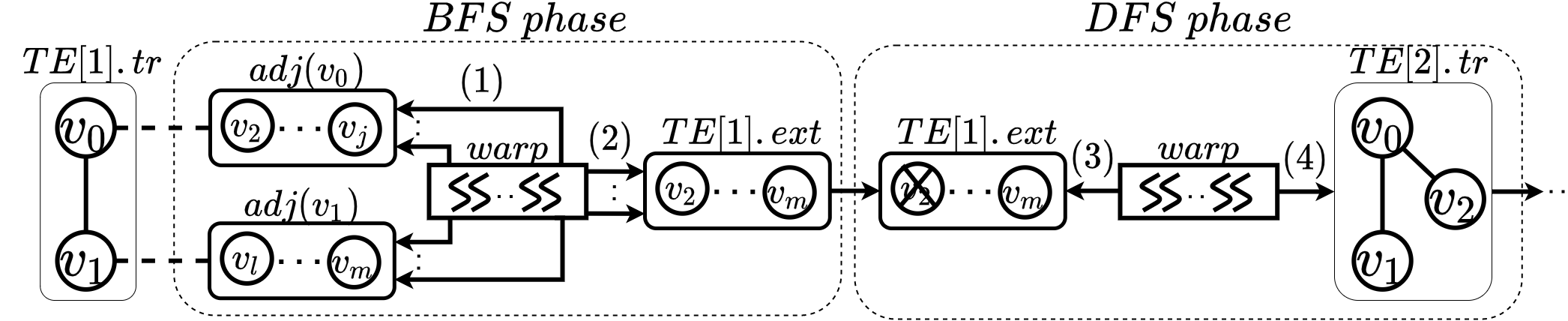}
  \caption{BFS and DFS steps.}
  \label{figure:dfs_wide_steps}
\end{subfigure}
\caption{DFS-wide subgraph exploration}
\label{figure:dfs_wide}
\end{figure}

We propose a novel DFS-wide subgraph exploration, which alternates between a BFS and a DFS phase to allow regular subgraph enumeration on GPU. Figure~\ref{figure:DFS_wide_overview} presents an overview of the DFS-wide exploration steps and Figure~\ref{figure:dfs_wide_steps} shows the operations performed in one iteration of BFS and DFS phases. $TE$ (\textit{Traversal Enumeration}) is an array that stores the intermediate states needed for DFS-wide. $TE.tr$ stores the vertex ids of the current traversal and $TE.ext$ stores the extensions generated through enumeration.

In Figure~\ref{figure:DFS_wide_overview}, the enumeration starts with a traversal $TE[i].tr$ and the BFS phase produces and stores the extensions efficiently in a contiguous array ($TE[i].ext$) which will be cached. The DFS receives the extensions and decides to move forward or backward in the enumeration, depending on the length of $tr$ and the extensions. Note that, in both forward and backward, the DFS phase will access extensions in a contiguous memory which is probably cached, improving memory efficiency. Enumeration proceeds alternating between BFS and DFS steps until the traversal reaches the target size. Assuming we want to enumerate a traversal $tr = \{v_0,v_1\}$, Figure~\ref{figure:dfs_wide_steps} details the operations performed in BFS and DFS phases in a single iteration of DFS-wide. In the BFS phase a warp visits the adjacency lists of vertices in current traversal (step 1), copies these vertices to extensions, and keeps only the unique extensions which are not in $tr$ (step 2). Once extensions are generated, the DFS phase starts by consuming a vertex ($v_2$) from extensions (step 3) and incrementing the current traversal (step 4).

The BFS phase is implemented by the warp-centric $Extend$ phase (described later in Section~\ref{subsection:extend}) of DuMato workflow, and the DFS phase is implemented by the warp-centric \emph{Move} phase (Algorithm~\ref{algorithm:move_phase}). \emph{Move} phase allows the warp to move forward/backward in the enumeration of a traversal. It receives $TE$ and a flag $genedges$ to indicate whether the edges of traversals should be generated throughout enumeration. The edges of traversals are useful in algorithms such as motif counting, where connectivity information is used to extract canonical representatives from subgraphs.
If necessary, the $Move$ phase generates edges gradually as enumeration progresses.
If the current traversal still has not reached the size limit and the current set of extensions is not empty (line 3), the warp moves forward in the enumeration by consuming an extension and extending the current traversal (lines 4,5). If the edges of traversal are needed, \textit{induce} function (line 6) is a SIMD step that reuses the edges of current traversal to produce the edges of the extended traversal. If either the current traversal has reached the size limit or the current extensions set is empty, the current traversal can not  be extended and the warp moves backward in the enumeration (line 7). In case the enumeration of current traversal finishes, the warp pulls a new traversal from a global queue (line 8). As all threads within a warp manipulate the same traversal and the main purpose of $Move$ is to update information about current traversal, it is mostly a SISD phase, and only \textit{induce} function that is costly is a SIMD step.

\begin{algorithm}[!hbt]
    \footnotesize
    \DontPrintSemicolon
        $\texttt{\textcolor{white}{SIMD}}\;void\;move(TE,\,genedges)\colon$\;
        $\texttt{\textcolor{black}{SISD}}\;\;\;\;extensions \leftarrow TE[TE.len-1].ext;$\;
        $\texttt{\textcolor{black}{SISD}}\;\;\;\;if(TE.len \neq k-1\;and\;extensions \neq \emptyset)\colon$\;
        $\texttt{\textcolor{black}{SISD}}\;\;\;\;\;\;\;ext \leftarrow pop(extensions);$\;
        $\texttt{\textcolor{black}{SISD}}\;\;\;\;\;\;\;TE[level+1].tr \leftarrow ext;$\;
        $\texttt{\textbf{\textcolor{black}{SIMD}}}\;\;\;\;\;\;\;if(genedges)\colon induce(TE);$\;
        $\texttt{\textcolor{black}{SISD}}\;\;\;\;else\colon TE.len--$\;
        $\texttt{\textcolor{black}{SISD}}\;\;\;\;if(TE.len = 0)\colon TE \leftarrow pull\;trav.\;from\;global\;queue;$\;
  \caption{Move primitive}
  \label{algorithm:move_phase}
\end{algorithm}

The worst-case space complexity of the DFS-wide exploration is $O(traversals \times max(G) \times k^2)$, where $traversals$ is the number of traversals processed in parallel, $max(G)$ is the maximum degree of the input graph, and $k$ is the length of explored subgraphs. All data structures are allocated in global memory and shared memory was set for caching, which is used in the BFS phase during the copy of adjacency lists to the extensions, as well as in the DFS phase to read an extension to move forward/backward. The cost for BFS subgraph exploration is $O(traversals \times max(G)^{k-1})$, which naturally leads to an exponential growth of memory demands as $k$ increases. The cost for DFS subgraph exploration is $O(traversals \times k)$, as the only intermediate state needed is the set of vertex ids of current traversal. Although DFS consumes less memory than DFS-wide, DFS-wide allows more regularity in execution and memory access pattern throughout subgraph enumeration.

Previous works have used DFS-like subgraph exploration for graph analytics algorithms such as PageRank and Connected Components~\cite{YXLL16,MBAMS18}. However, graph analytics algorithms rely on convergence parameters and the amount of intermediate states does not grows exponentially as in GPM algorithms, requiring a simpler system design. The next section explains our efficient \textit{warp-centric} strategies to improve memory coalescence and divergences.

\subsection{Efficient Warp-centric Filter-Process} \label{section:efficient_warp_centric_filter_process}

This section describes the warp-centric based design and
implementation of the DuMato phases of the filter-process workflow. Our goal with
this model is to minimize execution divergence in our
irregular algorithms, and to exploit the opportunities of parallelism and regular memory access enabled with
the DFS-wide strategy.
\subsubsection{Extend} \label{subsection:extend}

This phase is the BFS step that generates the neighbourhood extensions of a traversal $tr$ by visiting the adjacency lists of a specific range of vertices. This design is important to enable algorithms using the adjacency list of all vertices in the current traversal (e.g. motif counting) or only the adjacency list of a subset (e.g. clique counting).

Algorithm~\ref{algorithm:extend_primitive} shows our warp-centric implementation of the \emph{extend} phase. Every call to \emph{extend} returns a \emph{boolean} value to indicate whether its extensions had already been filled prior to the call, and this information is useful to avoid unnecessary calls to \emph{filter-compact}. Line 2 is an initial SISD phase, where all threads in the warp receive the array where extensions of current traversal are supposed to be written. In case the extensions have already been generated by previous calls to \emph{extend}, the function stops and returns \emph{false} (line 3). Lines 4-9 generate the extensions of current traversal by iterating the adjacency lists of vertices in $TE[start \cdots end]$.

 Threads in the warp receive the same vertex $id$ in the current traversal (line 4), whose adjacency will be visited in parallel. Each thread retrieves a different extension candidate $e$ by reading the adjacency of $id$ in parallel (line 5). As the adjacency list of a vertex is contiguous in global memory, this memory request is coalesced. Next, threads in the same warp work cooperatively to discover whether their extension candidates are valid for the current traversal. Threads compare their extension candidates to each vertex in current traversal to check whether they are already present in the current traversal (line 6). In this step, threads in the same warp execute in lockstep and compare their values to the same position $i$ in $TE[i].tr$, allowing broadcast of $TE[i].tr$ to all threads in the warp using one memory transaction. Next, threads compare their extension candidates to extensions already generated (line 7). In this step we also take advantage of lockstep execution and memory broadcasting, providing regular execution and reducing memory transactions. In case the extension candidates neither are in the current traversal nor in the extensions already generated, they are written to the current extensions in parallel through coalesced memory writes (lines 8-9). Note that all lines of \textit{extend} function are executed in lockstep by threads within a warp, minimizing divergences. Besides, each line also provides regular memory access patterns for all data structures, allowing memory coalescence and good cache locality. If new extensions were generated, the function returns $true$ (line 10).

\begin{algorithm}[!htb]
    \footnotesize
    \DontPrintSemicolon
    $\texttt{\textcolor{white}{SISD}}boolean\;extend(TE,\;start,\;end)\colon$\;
    $\texttt{\textcolor{black}{SISD}}\;\;\;\;len \leftarrow TE.len-1\;;\;extensions \leftarrow TE[len].ext;$\;
    $\texttt{\textcolor{black}{SISD}}\;\;\;\;if(extensions\;generated)\colon \mathbf{return} \; false;\;\;$\;
    $\texttt{\textcolor{black}{SISD}}\;\;\;\;for\;each(id \in TE[start \cdots end-1].tr)\colon\;\;$\;
    $\texttt{\textbf{\textcolor{black}{SIMD}}}\;\;\;\;\;\;for\;each(e \in adjacency(id))\colon\;\;$\;
    $\texttt{\textbf{\textcolor{black}{SIMD}}}\;\;\;\;\;\;\;\;\;inTraversal \leftarrow find\;e\;in\;TE[0\cdots len].tr;\;\;$\;
    $\texttt{\textbf{\textcolor{black}{SIMD}}}\;\;\;\;\;\;\;\;\;inExtensions \leftarrow find\;e\;in\;extensions;\;\;$\;
    $\texttt{\textbf{\textcolor{black}{SIMD}}}\;\;\;\;\;\;\;\;\;e \leftarrow \;!inTraversal\;\&\&\;!inExtensions \;?\;e\;:\;-1;\;\;$\;
    $\texttt{\textbf{\textcolor{black}{SIMD}}}\;\;\;\;\;\;\;\;\;write\;e\;to\;extensions;\;\;$\;
    $\texttt{\textcolor{black}{SISD}}\;\;\;\;\mathbf{return}\;true;\;\;$\;
  \caption{Extend primitive.}
  \label{algorithm:extend_primitive}
\end{algorithm}

\subsubsection{Filter} \label{subsection:filter}
Given a traversal $tr$, \textit{filter} (Algorithm~\ref{algorithm:filter}) phase iterates a set of extensions in parallel and invalidates those that do not meet a property $P$. Threads read consecutive extensions in parallel through coalesced memory accesses (line 3), call a function pointer $P$ to discover whether an extension satisfies the desired property (line 4), and invalidate (write $-1$ value) those which do not fulfill the property (line 4). The implementation of property functions $P$ is also warp-centric and uses DuMato primitives to access $TE$ data structure.

\begin{algorithm}[!htb]
    \footnotesize
    \DontPrintSemicolon
    $\texttt{\textbf{\textcolor{white}{SIMD}}}void\,filter(TE,\;P,\;args)\colon$\;
    $\texttt{\textcolor{black}{SISD}}\;\;\;extensions \leftarrow TE[TE.len-1].ext;\;\;$\;
    $\texttt{\textbf{\textcolor{black}{SIMD}}}\;\;\;for\;each(ext \in extensions)\colon\;\;$\;
    $\texttt{\textbf{\textcolor{black}{SIMD}}}\;\;\;\;\;\;if(P(TE,ext,args))\colon invalidate(extensions,i);\;\;$\;
    \caption{Filter primitive.}
    \label{algorithm:filter}
\end{algorithm}

\subsubsection{Compact} \label{subsection:compact}

It is an optional phase that accesses the extensions set of the current traversal and removes invalid positions, reducing its actual length. As seen in Algorithm~\ref{algorithm:filter}, a filter iterates over the entire set of extensions of the current traversal, even if some positions are invalid. By removing invalid positions, compaction reduces the costs of sequential filter calls. We provide an efficient warp-centric implementation of this function 
using intra-warp communication primitives such as \textit{ballot} and \textit{any\_sync}.

\subsubsection{Aggregate} \label{subsection:aggregate}
This phase is executed when a thread warp has derived traversals with $k$ vertices and, as discussed, it is in charge of producing the
actual GPM algorithm results (\textit{A} function of Eq.~\ref{eq:enumeration-function}). \textit{DuMato} provides three aggregation primitives: \textit{aggregate\_pattern}, \textit{aggregate\_counter} and \textit{aggregate\_store}, as defined in Table~\ref{tab:dumato-api}, which are explained below.

The \textit{aggregate\_pattern} is the most challenging primitive for 
implementation on GPUs. It is used when the output of the GPM 
algorithm relies on counting the occurrence of canonical representatives
(patterns) with $k$ vertices, such as motif counting. This is executed
in a warp basis such that each warp performs \textit{canonical relabeling}, converting each subgraph with $k$ vertices to its canonical representative and incrementing a counter. This is only possible due to our novel representation for
canonical representatives, 
which reduces the amount of memory required to store them. The solution for canonical relabeling relies on graph isomorphism, and GPM systems (including Pangolin~\cite{CDGP20}) perform 
it on CPU using tools such as Nauty~\cite{MP14}. To the best of our
knowledge, we are the first work to implement canonical 
relabeling on GPU.

\begin{figure}[!htb]
    \centering
    \includegraphics[width=1\linewidth]{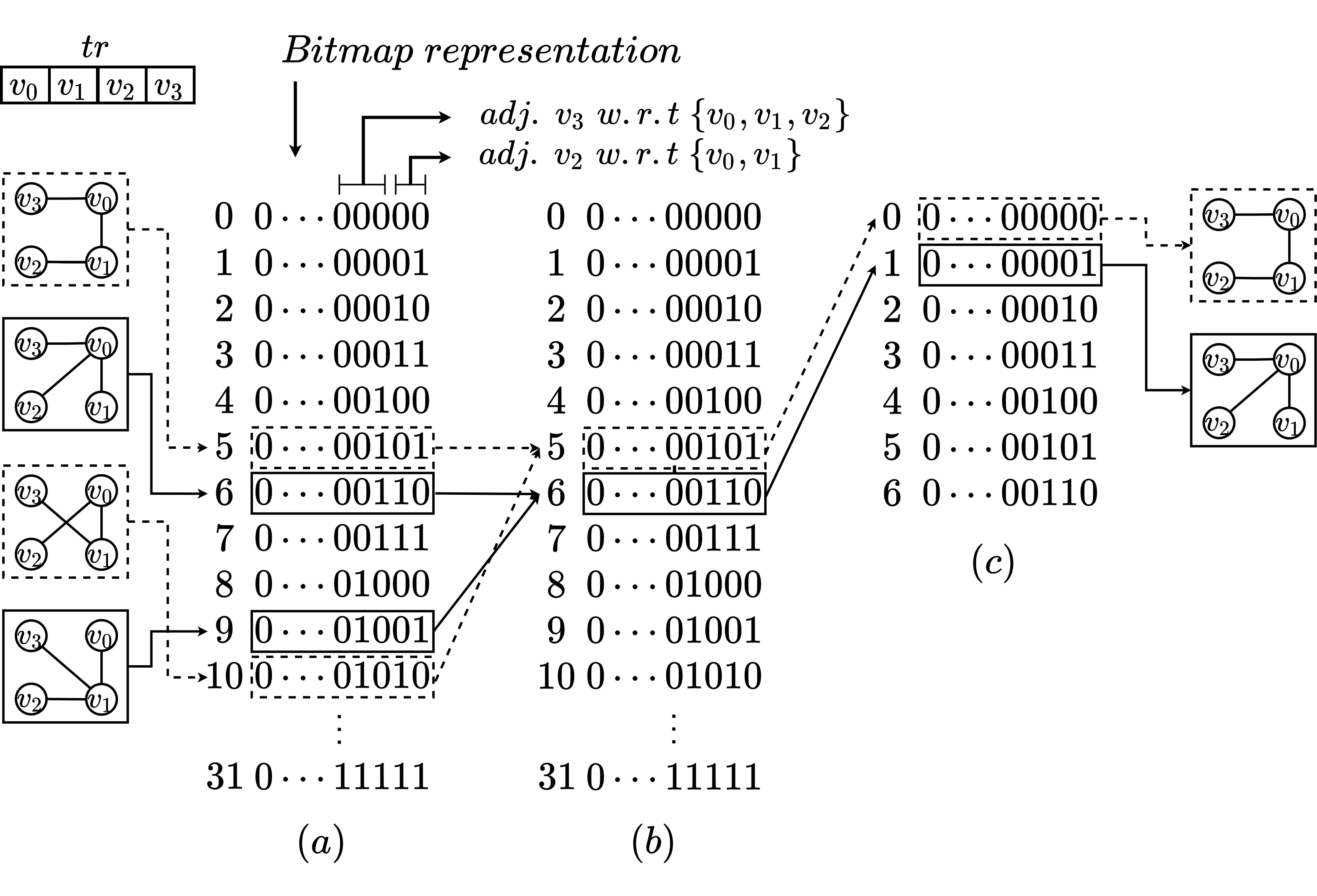}
    \caption{Canonical relabeling on GPU.}
    \label{figure:lcr}
\end{figure}

Figure~\ref{figure:lcr} depicts our strategy for canonical relabeling on GPU. We use a bitmap to store the edges of the traversal. For example, assuming $k=4$ and a traversal $tr$, we need $5$ bits to store the edges of a traversal. As we handle only connected traversals, $v_0$ is always connected to $v_1$ and this edge is not stored. The two least significant bits of the bitmap store the edges of $v_2$ with respect to $\{v_0,v_1\}$, and the next three bits store the edges of $v_3$ with respect to $\{v_0,v_1,v_2\}$ (the same reasoning may be applied to a subgraph with $k$ vertices). Using $5$ bits we can represent up to $32$ possible traversals, as seen in (a). Each possible traversal with $4$ vertices can be mapped to its canonical representative, shown in (b). As traversals often produce isomorphic subgraphs, different traversals may be mapped to the same canonical representative. The amount of canonical representatives is much smaller than the amount of possible traversals, as seen in (c), and the bitmap representation of canonical representatives can be relabeled to use consecutive bitmaps.

Our implementation creates a dictionary that receives a traversal $tr$ with $k$ vertices along with its edges encoded using the bitmap representation (a.k.a. an induced traversal) and converts $tr$ to a canonical representative that is in a contiguous range of positions (Figure~\ref{figure:lcr}).
This is done in two steps: in $(a)\to(b)$ traversal edges are mapped to non-contiguous representatives; and in $(b)\to(c)$ non-contiguous representatives are mapped to contiguous identifiers.
This conversion allows each warp to use local counters for canonical representatives using less memory, as no position in the array of counters is wasted. This dictionary is a pre-processed data structure, created once for a range of $k$ values, and that can be used in any dataset and in any application that requires canonical relabeling (e.g. frequent subgraph mining~\cite{EASK14} and subgraph matching~\cite{guo2020subgraphmatchinggpu}). DuMato provides this dictionary as an input file.

The \textit{aggregate\_counter} primitive is called when the desired results/output 
of the GPM algorithm is a pattern counting, such as in the clique counting
algorithm. Each warp produces its own counter (based on the length of the extensions for each traversal with $k-1$ vertices) to avoid inter-warp race conditions, and the global counting is produced with a reduction of the
warps counting afterwards, on CPU. This is a simple and computing inexpensive primitive. Primitive $aggregate\_store$ stores the explored subgraphs with $k$ vertices and can be used in algorithms such as subgraph querying, which lists all subgraphs that matches a pattern instead of producing counters. We create an array buffer that stores the connectivity bitmap of explored subgraphs with $k$ vertices as they are produced. DuMato then provides a producer-consumer environment using the CPU to consume the buffer asynchronously. 

\subsection{Warp-Level Load Balancing}\label{subsection:out_of_gpu_load_balancing}

\begin{figure}
    \centering
    \includegraphics[width=\linewidth]{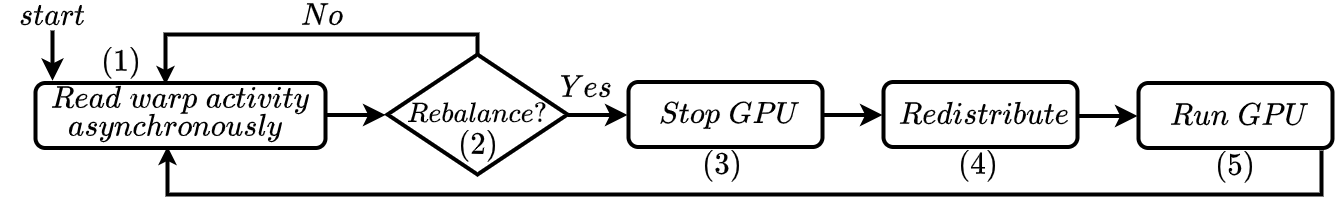}
    \caption{Warp-level load balancing.}
    \label{figure:load_balancing}
\end{figure}

The cost of enumerating distinct traversals may vary, which leads to load imbalance among warps. We propose 
an asynchronous workload redistribution scheme on CPU to mitigate this problem, depicted in Figure~\ref{figure:load_balancing}. Our strategy makes decisions based on the warp level activity information, and all steps are executed on CPU.

The CPU constantly and asynchronously
reads the warp activity information from the GPU to decide whether load
should be redistributed to improve GPU utilization (step 1). When CPU detects load balancing
is to be performed (step 2, \emph{rebalance}), the CPU informs the GPU by setting a flag and the warps
stop their execution in a consistent state (step 3). We propose a \textit{rebalance} condition such that, if the number of active warps is found to be lower than
a threshold, the workload balancing is carried out. When all warps stop, CPU copies $TE$ data structure, performs work redistribution (step 4, \emph{redistribute}) and runs kernel again on GPU (step 5). We propose a \textit{redistribute} algorithm that performs load balancing by separating warps in
\emph{donators} (those with multiple traversals) and \emph{idle} ones. While there
are \emph{idle} warps, a donator is selected (in a round-robin fashion) and 
an arbitrary traversal is migrated from it to the idle warp. We show the effectiveness of \emph{rebalance} and \emph{redistribute} strategies in our performance evaluation. As long as GPU is active, CPU inspects warp activity and perform work redistribution. 

\emph{Rebalance} and \emph{redistribute} steps in our scheme can be easily customized to implement other approaches. We highlight that, although it is possible to implement complex load balancing mechanisms using buffers to store jobs generated by warps~\cite{CT08}, our CPU-only strategy mitigates synchronization overheads from GPU and more resources can be allocated for subgraph enumeration.

\subsection{Programming with DuMato} \label{section:programming_dumato}

DuMato's workflow (Fig.~\ref{figure:execution_workflow}) is able to represent any GPM algorithm relying on the enumeration of induced subgraphs. Table~\ref{tab:dumato-api} shows the programming interface that can be used to create algorithms in DuMato using our efficient strategies. The functions receive as parameter a data structure holding runtime information about the active traversal and extension arrays ($TE$, detailed in Fig.~\ref{figure:dfs_wide}) along with additional parameters.

\begin{table}[!htb]
    \centering
    \footnotesize
    \begin{tabular}{l l l l}
        \toprule
        \textbf{Functions} &  \textbf{Phase} & \textbf{Scope}
        \vspace{1pt}\\\hline
        \texttt{[CT]} \textit{control(TE)} & Control & \multirow{2}{*}{Algorithm-independent}
        \\
        \texttt{[MV]} \textit{move(TE, genedges)} & Move &
        \\\hline
        \texttt{[EX]} \textit{extend(TE, begin, size)} & Extend & \multirow{6}{*}{Algorithm-specific}
        \vspace{3pt} \\
        \texttt{[FL]} \textit{filter(TE, P, args)} & Filter
        \vspace{3pt} & &  \\
        \texttt{[CP]} \textit{compact(TE)} & Compact & 
        \vspace{3pt} \\
        \texttt{[A1]} \textit{aggregate\_counter(TE)} &\multirow{3}{*}{Aggregate} & 
        \\
        \texttt{[A2]} \textit{aggregate\_pattern(TE)} & & 
        \\
        \texttt{[A3]} \textit{aggregate\_store(TE)} & & 
        \\
        \bottomrule
    \end{tabular}
    \caption{DuMato API.}
    \label{tab:dumato-api}
\end{table}

\emph{Control} and \emph{Move} phases are responsible for keeping the workflow active while there are unprocessed traversals in the search space. Because this loop-based exploration is common to most GPM algorithms searching for multiple subgraphs, we say that these phases are independent of algorithm semantics. Functions \texttt{[CT]} and \texttt{[MV]} implement these two phases. \texttt{[CT]} allows the underlying runtime to check the termination conditions of the execution.
\texttt{[MV]} implements the traversal order of exploration and receives an additional parameter $genedges$ that determines whether the edges of the current traversal should be generated.

\emph{Extend}, \emph{Filter}, \emph{Compact}, and \emph{Aggregate} phases enables a straightforward and efficient representation of application-specific semantics on GPUs.
Function \texttt{[EX]} implements the \emph{Extend} phase and generates the extensions array by fetching the neighborhood of vertices in the traversal at positions in range $[begin,size)$. This can be used to generate extensions using alternative strategies that may be more effective to explore subgraphs having patterns known apriori~\cite{danisch2018kclist,JMV20}.
\texttt{[FL]} implements the \emph{Filter} phase and allows invalidating extensions that do not satisfy a user-defined property. The input to this call is a pointer to a function and its arguments (property $P$ and $args$, respectively) that is applied to each extension to maintain only valid ones. This interface can be used to design custom subgraph filters of extensions based on canonical candidate generation~\cite{TFSSZA15}, density~\cite{liu2008quasicliques}, subgraph matching~\cite{guo2020subgraphmatchinggpu}, among others.
\texttt{[CP]} implements the \emph{Compact} phase and can be applied between consecutive \texttt{[FL]} calls to compact the extensions array. This interface allows a fine-grained control over the underlying GPU memory organization, which can be useful to speedup memory access when it is possible to infer the selective potential of filters from application-specific semantics~\cite{eppstein2013maximalcliques,danisch2018kclist}.
\texttt{[A1]}, \texttt{[A2]}, and \texttt{[A3]} implement the \emph{Aggregate} phase:
\texttt{[A1]} counts the number of valid extensions in the array of extensions;
\texttt{[A2]} counts the number of traversals per pattern; and
\texttt{[A3]} allows buffering of traversals for custom semantics and further downstream processing. These can be used, for instance, for subgraph counting~\cite{danisch2018kclist} and scoring~\cite{hooi2020telltail}.

Algorithms are implemented in a loop that processes new traversals until the termination condition is reached. After each loop iteration, DuMato moves the exploration to a new traversal as a preparation to the next iteration. This is common to most GPM algorithms and can be observed in lines 11 and 20 of Algorithm~\ref{algorithm:dumato_clique_motifs}, which presents the implementation of two representative GPM algorithms using DuMato API: \emph{clique counting} and \emph{motif counting}.
Bold lines marked with \usericon~ represent algorithm-specific semantics that uses DuMato's API and, consequently, new algorithms with new extend, filtering, and aggregation demands may be implemented by replacing those lines. 

\begin{algorithm}[!htb]
\begin{multicols}{2}
    \scriptsize
    \DontPrintSemicolon
    $\,\,\,void\; clique\_counting (TE)\colon$\;
    $\,\,\,\,\,while(control(TE))\colon$\;
    \usericon $\,\,\,\,\,\,\,\mathbf{if(extend(TE,0,1))\colon}$\,\,\,\;
    \usericon $\,\,\,\,\,\,\,\,\,\,\mathbf{u \gets TE[TE.len-1].id;}$\,\,\,\;
    \usericon $\,\,\,\,\,\,\,\,\,\,\mathbf{filter(TE,\,\&lower\,,\,[u]);}$\,\,\,\;
    \usericon $\,\,\,\,\,\,\,\,\,\,\mathbf{compact(TE);}$\,\,\,\;
    \usericon $\,\,\,\,\,\,\,\,\,\,\mathbf{filter(TE, \&is\_clique, []);}$\,\,\,\;
    $\,\,\,\,\,\,\,\,\,\,if(TE.len = k-1)\colon$\;
    \usericon $\,\,\,\,\,\,\,\,\,\,\mathbf{aggregate\_counter(TE);}$\,\,\,\;
    $\,\,\,\,\,\,\,\,\,\,move(TE,\,false);$\;
    \vfill\null
    \columnbreak
    $\,\,\,void\;motif\_counting(TE)\colon$\;
    $\,\,\,\,\,\,while(control(TE))\colon$\;
    \usericon $\,\,\,\,\,\,\mathbf{if(extend(TE,0,TE.len))}\colon$\,\,\,\;
    \usericon $\;\;\;\;\;\;\mathbf{filter(TE,\&canonical,[]);}$\,\,\,
    \;
    $\,\,\,\,\,\,\,\,\,if(TE.len = k-1)\colon$\;
    \usericon $\;\;\;\;\;\;\mathbf{aggregate\_pattern(TE);}$\,\,\,\;
    $\,\,\,\,\,\,\,\,\,move(TE,\,true);$\;
  \end{multicols}
  \caption{Clique and motif counting algorithms.}
  \label{algorithm:dumato_clique_motifs}
\end{algorithm}

\emph{Clique counting.} 
A clique of size $k$ is a graph $C$ with $k$ vertices such that, for every $v_i \in V(C)$ and $v_j \in V(C)$, $(v_i,v_j) \in E(C)$. Given a graph $G$, the clique counting problem seeks to count the number of cliques with $k$ vertices within $G$.
Clique counting represents algorithms whose goal is searching for subgraphs with the same pattern.
Because a clique extension must be adjacent with every vertex in the traversal, the \emph{Extend} phase consists of generating the array of extensions from the neighbors of a single vertex in the traversal. \texttt{[EX]} call in line 3 of Algorithm~\ref{algorithm:dumato_clique_motifs} implements this idea by indicating that the current extensions should be obtained from the neighbors of the first vertex in the traversal
(represented by the range $[0,1)$).
Given this set of extensions, \texttt{[FL]} is used in line 5 to invalidate non-canonical candidates (extensions lower than the last vertex), \texttt{[CP]} is used in line 6 to reorganize the extensions array by compaction, and \texttt{[FL]} is used again in line 7 to remove extensions that do not generate cliques.
The custom procedure $is\_clique$ ensures that valid extensions are connected to all vertices in the traversal. Both $lower$ and $is\_clique$ are simple functions that must return $true$ or $false$ given a traversal and one of its extensions.
Finally, if the traversal reaches $k-1$ vertices then traversals with $k$ vertices may be aggregated with \texttt{[A1]}, which accumulates the length of the array of extensions in a counter.

\emph{Motif counting.}
A motif of size $k$ is a canonical representative subgraph containing $k$ vertices. The motif counting problem seeks to count the number of each motif of size $k$ in a graph $G$.
Motif counting represents algorithms whose target is searching for subgraphs of multiple patterns.
Because this problem requires visiting all induced subgraphs of size $k$, the \texttt{[EX]} call in line 15 indicates that the adjacency of each vertex in the traversal must be considered to produce the extensions array (i.e. all traversal vertices in range $´[0,TE.len)$). In line 17 the algorithm calls \texttt{[FL]}
to invalidate extensions that combined with the traversal do not represent canonical candidates.
Custom function $is\_canonical$ can be implemented using standard \emph{canonical filtering algorithms}~\cite{TFSSZA15}.
Finally, a \texttt{[A2]} call extracts the canonical representative (pattern) from traversals combined with last level extensions to increment the respective pattern-specific counters (line 19).

\section{Evaluation}\label{section:evaluation}
This section presents the DuMato performance evaluation. We employ the implementations of \textit{clique counting} and \textit{motif counting} algorithms. The algorithms represent two important categories in GPM processing: exploration of subgraphs sharing a single canonical representative (Clique counting) and exploration of subgraphs ranging multiple canonical representatives (Motif counting).
The attributes of five real-world datasets used in our experiments are presented in Table~\ref{table:datasets}. CPU experiments were conducted on an Amazon AWS machine with 16 vCPUs optimized for CPU computing (\emph{C5a.4xlarge}), 32GB of RAM and Ubuntu 22.04. GPU experiments concerning execution time were conducted on an Amazon AWS machine with one NVIDIA Testa V1OO (\emph{p3.2xlarge}) with 32 Gb and CUDA 11. GPU profiling experiments were conducted on a local machine with NVIDIA TITAN V with 12GB and CUDA 10.1. The time limit adopted for each execution was 24 hours. After a theoretical occupancy analysis and an empirical evaluation, the configuration of the experiments was set to 172,032 threads for all datasets. This amount of threads was enough to keep SMs busy without overloading GPU registers. For the motif counting we do not present the results for LiveJournal graph because the executions exceed our 24 hours limit even for small subgraph sizes ($k>4$).

\begin{table}[!hbt]
\centering
\footnotesize
\setlength\tabcolsep{1pt} 
\begin{tabular}{|c|r|r|r|c|r|}
\hline
\textbf{Dataset} & \textbf{V(G)} & \textbf{E(G)} & \textbf{Avg. Deg.} & \textbf{Density} & \textbf{Max. Deg.} \\ \hline
Citeseer~\cite{EASK14}         & $3.2$K          & $4.5$K          & $2.77$                 & $8.51 \times 10^{-4}$     & $99$                   \\ \hline
ca-AstroPh~\cite{LKF07}       & $18.7$K         & $198.1$K        & $21.10$                & $1.12 \times 10^{-3}$     & $504$                  \\ \hline
Mico~\cite{EASK14}             & $96.6$K         & $1.08$M       & $22.35$                & $2.31 \times 10^{-4}$     & $1359$                 \\ \hline
com-DBLP~\cite{YL12}             & $317$K        & $1.04$M       & $6.62$                 & $2.08 \times 10^{-5}$     & $343$                  \\ \hline
com-LiveJournal~\cite{YL12}  & $3.9$M       & $34.6$M      & $17.35$                 & $4.34 \times 10^{-6}$     &                      $14815$ \\ \hline
\end{tabular}
\caption{Graphs used for evaluation.} \label{table:datasets}
\end{table}

\subsection{Impact of Optimizations}
\label{sec:impact-optimizations}
This section evaluates the efficiency of our optimization strategies using three implementations of clique and motif counting (all implemented with DuMato API):
\textit{DM\_DFS} (\textit{\textbf{D}u\textbf{M}ato \textbf{D}epth-\textbf{F}irst \textbf{S}earch}), in which each GPU thread receives a traversal $tr$ and calculates $E(G,\allowbreak tr,\allowbreak k,\allowbreak P)$ using DFS exploration;
\textit{DM\_WC} (\textit{\textbf{D}u\textbf{M}ato \textbf{W}arp-\textbf{C}entric}), in which each warp receives a traversal and enumerates it using DFS-wide and the warp-centric design, but with load balancing disabled;
\textit{DM\_OPT} (\textit{\textbf{D}u\textbf{M}ato \textbf{Opt}imized}), which is DM\_WC with load balancing enabled. Table~\ref{tab:optimizations-performance} shows the execution times for the three
versions of both algorithms as the length of the subgraphs mined ($k$) is varied. Cells containing ``-" refer to experiments that have not finished within 24 hours.

\begin{table}[!htb]
\centering
\tiny
\renewcommand{\arraystretch}{0.9}
\setlength\tabcolsep{6pt}
\begin{tabular}{c| ll rrrrr}

\toprule
    & 
    & Impl.
    & $k=3$
    & $k=4$
    & $k=5$
    & $k=6$
    & $k=7$
    \\
    \midrule
    
\multirow{16.6}{*}{Clique}
    & \multirow{3}{*}{Citeseer}
    & \texttt{DM\_DFS}
    & $0.01$ & $0.01$ & $0.01$ & $0.01$ & $\emptyset$
    \\
    &
    & \texttt{DM\_WC}
    & $0.01$ & $0.01$ & $0.02$ & $0.02$ & $\emptyset$
    \\
    &
    & \texttt{DM\_OPT}
    & $0.01$ & $0.01$ & $0.02$ & $0.03$ & $\emptyset$
    \\
    \cmidrule(lr){2-8}
    
    & \multirow{3}{*}{ca-AstroPh}
    & \texttt{DM\_DFS}
    & $0.23$ & $4.75$ & $51.43$ & $430.11$ & $44.78$K
    \\
    &
    & \texttt{DM\_WC}
    & $0.03$ & $0.36$ & $3.50$ & $28.98$ & $221.75$
    \\
    &
    & \texttt{DM\_OPT}
    & $0.13$ & $0.28$ & $0.67$ & $2.37$ & $11.46$
    \\
    \cmidrule(lr){2-8}

    & \multirow{3}{*}{Mico}
    & \texttt{DM\_DFS}
    & $3.28$ & $267.32$ & $19.67$K & - & -
    \\
    &
    & \texttt{DM\_WC}
    & $0.26$ & $12.62$ & $593.94$ & $26.31$K & -
    \\
    &
    & \texttt{DM\_OPT}
    & $0.33$ & $1.93$ & $51.98$ & $1.75$K & -
    \\
    \cmidrule(lr){2-8}
    
    & \multirow{3}{*}{DBLP}
    & \texttt{DM\_DFS}
    & $0.16$ & $4.04$ & $134.13$ & $3.64$K & -
    \\
    &
    & \texttt{DM\_WC}
    & $0.03$ & $0.33$ & $8.04$ & $232.96$ & $5.63$K
    \\
    &
    & \texttt{DM\_OPT}
    & $0.13$ & $0.28$ & $1.01$ & $7.14$ & $96.22$
    \\
    \cmidrule(lr){2-8}

clique - livejournal
    & \multirow{3}{*}{LiveJournal}
    & \texttt{DM\_DFS}
    & $337.85$ & $6.65$K & - & - & -
    \\
    &
    & \texttt{DM\_WC}
    & $16.83$ & $260.25$ & $6.77$K & - & -
    \\
    &
    & \texttt{DM\_OPT}
    & $4.30$ & $49.82$ & $897.25$ & $38.50$K & -
    \\
    \midrule
    
\multirow{12}{*}{Motifs}
    & \multirow{3}{*}{Citeseer}
    & \texttt{DM\_DFS}
    & $0.01$ & $0.49$ & $10.84$ & $232.11$ & $6.11$K
    \\
    &
    & \texttt{DM\_WC}
    & $0.01$ & $0.06$ & $1.26$ & $25.90$ & $457.17$
    \\
    &
    & \texttt{DM\_OPT}
    & $0.11$ & $0.11$ & $0.23$ & $0.68$ & $8.27$
    \\
    \cmidrule(lr){2-8}
    
    & \multirow{3}{*}{ca-AstroPh}
    & \texttt{DM\_DFS}
    & $1.59$ & $555.64$ & - & - & -
    \\
    &
    & \texttt{DM\_WC}
    & $0.09$ & $20.26$ & $5.28$K & - & -
    \\
    &
    & \texttt{DM\_OPT}
    & $0.13$ & $1.78$ & $149.43$ & $28.14$K & -
    \\
    \cmidrule(lr){2-8}

    & \multirow{3}{*}{Mico}
    & \texttt{DM\_DFS}
    & $20.58$ & $13.90$K & - & - & -
    \\
    &
    & \texttt{DM\_WC}
    & $0.95$ & $597.66$ & - & - & -
    \\
    &
    & \texttt{DM\_OPT}
    & $0.46$ & $33.44$ & $10.56$K & - & -
    \\
    \cmidrule(lr){2-8}
    
    & \multirow{3}{*}{DBLP}
    & \texttt{DM\_DFS}
    & $0.96$ & $178.69$ & $26.69$K & - & -
    \\
    &
    & \texttt{DM\_WC}
    & $0.08$ & $8.98$ & $1.35$K & - & -
    \\
    &
    & \texttt{DM\_OPT}
    & $0.14$ & $1.05$ & $38.07$ & $2.95$K & -
    \\
    \bottomrule
    \multicolumn{8}{c}{$\emptyset$: no valid subgraphs}
\end{tabular}
\caption{Optimizations performance. Execution time (seconds) for three implementations of algorithms using DuMato.}
\label{tab:optimizations-performance}
\end{table}

\subsubsection{Gains Due to Warp-centric DFS-wide}
The DFS version consumes a small amount of memory, but each thread within a warp has its own execution path, leading to an irregular execution and a worse memory access pattern. The DM\_WC version increases both memory efficiency and parallelism regularity, achieving speedups up to 33x (Clique app, Mico dataset and $k = 5$) compared to DM\_DFS, and showing the efficiency of memory coalescence and divergence reduction. We also observe that DM\_WC does not perform better than DM\_DFS whenever the amount of parallel work is insufficient, as in clique counting in \emph{Citeseer}, which contains few cliques.

To understand the effects of our exploration and optimization strategies at hardware level, Table~\ref{table:profiling} shows the improvements of \emph{DM\_WC} over \emph{DM\_DFS} using execution and memory metrics collected from CUDA NVProf profiling tool~\cite{toolkit_doc}. We present the results using DBLP dataset for $k$ up to $4$ (GPU profiling is much slower than standard runs). Metrics are divided into two categories: (i) Execution, which measures the efficient use of GPU execution model and parallelism and (ii) Memory, which quantifies the use of the memory hierarchy. For execution, we chose the metric \emph{inst\_per\_warp}, which calculates the average number of instructions executed by each warp. The more regular the execution is, the less divergent instructions are issued and warps require less instructions. For memory, we chose the metric \emph{gld\_transactions}, which measures the total amount of load transactions requested to global memory. The more coalesced is the memory access pattern, the less transactions are needed to service memory requests. In our experiments, we observed that the other metrics were consistent with these two representative choices.

\begin{table}[]
\centering
\scriptsize
\renewcommand{\arraystretch}{1.1}
\setlength\tabcolsep{3pt} 
\begin{tabular}{c|ccccccc}
\toprule

\multirow{2}{*}{App.} 
& 
\multirow{2}{*}{$k$} 
& 
\multicolumn{3}{c}{Memory (load transactions)} 
& 
\multicolumn{3}{c}{Execution (inst. per warp)}
\\

&

&
\texttt{DM\_DFS}
&
\texttt{DM\_WC}
&
Improvement
&
\texttt{DM\_DFS}
&
\texttt{DM\_WC}
&
Improvement
\\
\midrule

\multirow{2}{*}{Clique}
& 
$3$
&
$618.1$M
&
$212.7$M
&
$2.9\times$
&
$3.3$M
&
$876.6$K
& 
$3.8\times$
\\

& 
$4$
&
$6.7$B
&
$852.4$M
&
$7.9\times$
&
$50.5$M
&
$5.1$M
& 
$9.9\times$
\\
\midrule

\multirow{2}{*}{Motifs}
& 
$3$
&
$3.3$B
&
$597.0$M
&
$5.53\times$
&
$17.5$M
&
$2.6$M
& 
7.36$\times$
\\

& 
$4$
&
$134.7$B
&
$22.8$B
&
$5.90\times$
&
$1.9$B
&
$143.2$M
& 
$13.3\times$
\\
\bottomrule
\end{tabular}
\caption{Improvements of \texttt{DM\_WC} over \texttt{DM\_DFS}.}
\label{table:profiling}
\end{table}

\emph{Execution metrics}: The Warp-Centric DFS-Wide exploration results in natural lockstep implementation, which fits better GPU execution model and allows all threads within a warp to execute the same instruction more often to minimize divergence.
This reduces the total number of 
instructions per warp for the $DM\_WC$ version, as they execute mostly in lockstep and all threads in the warp tend to execute the same instruction. This regularity is confirmed by the execution metrics, with improvements ranging from 3.8x and 13.3x, confirming that our warp-centric design provides more regular execution.

\emph{Memory metrics:} The Warp-Centric DFS-Wide exploration with its regular lockstep execution allowed threads to perform memory requests together using coalesced requests. Therefore, our $DM\_WC$ version reduces the total amount of memory transactions. This reduction is confirmed by the memory metric, with improvements ranging from 2.90x to 7.92x, enhancing that our memory optimizations reduce wasted bandwidth and improve memory efficiency.

\subsubsection{Improvements with Load Balancing} \label{subsection:hybrid_load_balancing}

In order to define the adequate load balancing threshold, we conducted a sensitivity analysis (not shown due to space constraints) varying the amount of threads and the threshold used for rebalancing. We found that $172,032$ threads were enough to provide massive parallelism without overloading register allocation. We also found that, for this amount of threads, the optimum load balancing threshold was $40\%$ for clique counting and $10\%$ for motif counting. As clique counting prunes the search space, load imbalance occurs earlier than in motif counting, requiring a larger value of threshold to avoid excessive calls to the load balancing layer.

Table~\ref{tab:optimizations-performance} shows that the DM\_OPT version attained speedups of up to 65$\times$ compared to WC (Motifs app, \emph{Citeseer} dataset and $k = 8$). As the size of enumerated subgraphs increases, work skewness is intensified because most subgraphs are extracted from denser regions of the graph associated with increasingly fewer vertices and, at this point, load balancing becomes more effective. Hence, DM\_OPT allowed the exploration of larger subgraphs for all datasets. We also observe that DM\_OPT is not always beneficial compared to DM\_WC, especially for $k \leq 4$ in small datasets. In particular, whenever the amount of work is insufficient to exhibit a substantial imbalance or to payoff the overhead of redistributing the load, DM\_WC outperforms DM\_OPT.

\subsection{Comparison to Other GPM Systems} \label{sec:comparative-performance}

This section compares our optimal DuMato GPU implementations (DM\_OPT) against three representative state-of-the-art GPM systems: Pangolin~\cite{CDGP20} GPM system designed for GPU, and Fractal~\cite{DTGMP19} and Peregrine~\cite{JMV20} parallel CPU machines. Table~\ref{tab:comparative-performance} also shows the results. DuMato is more scalable and able to explore larger subgraphs than all  baselines within the same time limit, exploring subgraphs of up to 12 vertices. To the best of our knowledge, this length of explored subgraphs has not been accomplished by any other GPM system searching for exact outputs.

\begin{table}[!htb]
\centering
\tiny
\renewcommand{\arraystretch}{0.9}
\setlength\tabcolsep{1.2pt} 
\begin{tabular}{c| cl rrrrrrrrrr}

\toprule
    & 
    & System
    & $k=3$
    & $k=4$
    & $k=5$
    & $k=6$
    & $k=7$
    & $k=8$
    & $k=9$
    & $k=10$
    & $k=11$
    & $k=12$
    \\
    \midrule
    
\multirow{24}{*}{\rotatebox{90}{Clique}}
    & \multirow{4}{*}{\rotatebox{90}{Citeseer}}
    & \texttt{\textbf{DM}}
    & $0.01$ & $0.01$ & $0.01$ & $0.01$ & $\emptyset$ & $\emptyset$ & $\emptyset$ & $\emptyset$ & $\emptyset$ & $\emptyset$
    \\
    &
    & \texttt{FRA}
    & $4.84$ & $4.83$ & $4.75$ & $4.81$ & $\emptyset$ & $\emptyset$ & $\emptyset$ & $\emptyset$ & $\emptyset$ & $\emptyset$
    \\
    &
    & \texttt{PER}
    & $0.01$ & $0.03$ & $0.02$ & $0.02$ & $\emptyset$ & $\emptyset$ & $\emptyset$ & $\emptyset$ & $\emptyset$ & $\emptyset$
    \\
    &
    & \texttt{PAN}
    & $0.01$ & $0.01$ & $0.01$ & $0.01$ & $\emptyset$ & $\emptyset$ & $\emptyset$ & $\emptyset$ & $\emptyset$ & $\emptyset$
    \\
    \cmidrule(lr){2-13}
    
    & \multirow{4}{*}{\rotatebox{90}{ca-AstroPh}}
    & \texttt{\textbf{DM}}
    & $0.13$ & $0.28$ & $0.67$ & $2.37$ & $11.46$ & $57.55$ & $297.89$ & $1.47$K & $6.73$K & $28.14$K
    \\
    &
    & \texttt{FRA}
    & $8.17$ & $9.75$ & $15.89$ & $78.09$ & $439.16$ & $2.30$K & $12.89$K & $57.02$K & - & -
    \\
    &
    & \texttt{PER}
    & $0.01$ & $0.10$ & $0.83$ & $6.38$ & $43.56$ & $272.42$ & $1.55$K & $7.93$K & $36.26$K & -
    \\
    &
    & \texttt{PAN}
    & $0.01$ & $0.01$ & $0.02$ & $0.11$ & $0.61$ & OOM & OOM & OOM & OOM & -
    \\
    \cmidrule(lr){2-13}
 
    & \multirow{4}{*}{\rotatebox{90}{Mico}}
    & \texttt{\textbf{DM}}
    & $0.42$ & $3.89$ & $57.49$ & $2.22$K & - & - & - & - & - & -
    \\
    &
    & \texttt{FRA}
    & $14.17$ & $48.53$ & $1.44$K & $56.72$K & - & - & - & - & - & -
    \\
    &
    & \texttt{PER}
    & $0.09$ & $1.81$ &	$82.67$ & $3.66$K & - & - & - & - & - & -
    \\
    &
    & \texttt{PAN}
    & $0.01$ & $0.05$ & $2.93$ & OOM & - & - & - & - & - & -
    \\
    \cmidrule(lr){2-13}
    
    & \multirow{4}{*}{\rotatebox{90}{DBLP}}
    & \texttt{\textbf{DM}}
    & $0.13$ & $0.28$ & $1.01$ & $7.14$ & $96.22$ & $1.45$K & $20.86$K & - & - & -
    \\
    &
    & \texttt{FRA}
    & $13.44$ & $14.32$ & $22.72$ & $186.97$ & $2.52$K & $35.51$K & - & - & - & -
    \\
    &
    & \texttt{PER}
    & $0.11$ & $0.16$ & $1.36$ & $25.92$ & $531.88$ & $9.35$K & - & - & - & -
    \\
    &
    & \texttt{PAN}
    & $0.01$ & $0.01$ & $0.03$ & $0.50$ & OOM & OOM & OOM & OOM & OOM & OOM
    \\
    \cmidrule(lr){2-13}
 
    & \multirow{4}{*}{\rotatebox{90}{LiveJournal}}
    & \texttt{\textbf{DM}}
    & $4.30$ & $49.82$ & $897.25$ & $38.50$K & - & - & - & - & -  & -
    \\
    &
    & \texttt{FRA}
    & $394.85$ & $901.05$ & $16.06$K & - & - & - & - & - & - & -
    \\
    &
    & \texttt{PER}
    & $3.91$ & $26.66$ & $1.06$K & $64.74$K & - & - & - & - & - & -
    \\
    &
    & \texttt{PAN}
    & $0.01$ & $0.53$ & OOM & OOM & OOM & OOM & OOM & OOM & OOM & OOM
    \\
    \\
    \midrule
     
\multirow{20}{*}{\rotatebox{90}{Motifs}}
    & \multirow{4}{*}{\rotatebox{90}{Citeseer}}
    & \texttt{\textbf{DM}}
    & $0.11$ & $0.11$ & $0.23$ & $0.68$ & $8.27$ & $157.97$ & - & - & - & -
    \\
    &
    & \texttt{FRA}
    & $5.17$ & $5.20$ & $5.69$ & $12.44$ & $163.48$ & - & - & - & - & -
    \\
    &
    & \texttt{PER}
    & $0.01$ & $0.01$ & $0.05$ & $3.47$ & $537.66$ & - & - & - & - & -
    \\
    &
    & \texttt{PAN}
    & $0.01$ & $0.01$ & INC & OOM & OOM & OOM & OOM & OOM & OOM & OOM
    \\
    \cmidrule(lr){2-13}
     
    & \multirow{4}{*}{\rotatebox{90}{ca-AstroPh}}
    & \texttt{\textbf{DM}}
    & $0.13$ & $1.78$ & $149.43$ & $28.14$K & - & - & - & - & - & -
    \\
    &
    & \texttt{FRA}
    & $9.13$ & $435.64$ & $4.72$K & - & - & - & - & - & - & -
    \\
    &
    & \texttt{PER}
    & $0.01$ & $0.57$ & $132.90$ & $52.80$K & - & - & - & - & - & -
    \\
    &
    & \texttt{PAN}
    & $0.01$ & $0.21$ & INC & OOM & OOM & OOM & OOM & OOM & OOM & OOM
    \\
    \cmidrule(lr){2-13}
 
    & \multirow{4}{*}{\rotatebox{90}{Mico}}
    & \texttt{\textbf{DM}}
    & $0.46$ & $33.44$ & $10.56$K & - & - & - & - & - & - & -
    \\
    &
    & \texttt{FRA}
    & $16.43$ & $474.46$ & - & - & - & - & - & - & - & -
    \\
    &
    & \texttt{PER}
    & $0.06$ & $6.57$ & $7.92$K & - & - & - & - & - & - & -
    \\
    &
    & \texttt{PAN}
    & $0.01$ & $3.31$ & OOM & OOM & OOM & OOM & OOM & OOM & OOM & OOM
    \\
    \cmidrule(lr){2-13}
     
    & \multirow{4}{*}{\rotatebox{90}{DBLP}}
    & \texttt{\textbf{DM}}
    & $0.14$ & $1.05$ & $38.07$ & $2.95$K & - & - & - & - & - & -
    \\
    &
    & \texttt{FRA}
    & $14.33$ & $37.62$ & $1.43$K & - & - & - & - & - & - & -
    \\
    &
    & \texttt{PER}
    & $0.07$ & $0.95$ & $78.59$ & $50.95$K & - & - & - & - & - & -
    \\
    &
    & \texttt{PAN}
    & $0.01$ & $0.17$ & INC & OOM & OOM & OOM & OOM & OOM & OOM & OOM
    \\
    \bottomrule

\multicolumn{13}{c}{\texttt{DM}: DuMato (this work); \texttt{FRA}: Fractal; \texttt{PER}: Peregrine; \texttt{PAN}: Pangolin}
\\
\multicolumn{13}{c}{\textit{OOM}: out-of-memory; \textit{INC}: incomplete results; $\emptyset$: no valid subgraphs}
    
\end{tabular}
\caption{Comparative performance. Execution time (seconds) of DuMato and baselines (GPU and CPU).}
\label{tab:comparative-performance}
\end{table}

Pangolin clearly suffers from scalability issues. Although it achieves good 
performance for small datasets and small enumerated subgraphs, it usually 
runs out of memory when the length of explored subgraphs is close to 5 vertices, limiting its applicability and the discoveries of GPM algorithms. As compared to Fractal, we obtain significant speedups in all executions  with gains ranging from 17$\times$ to 103$\times$. In general, as length of explored subgraphs increases, the processing cost is higher and DuMato can take more advantage of GPU's massive parallel processing and to achieve better gains.

Regarding Peregrine, DuMato is competitive for small values of explored 
subgraphs (up to 5 vertices), and shows speedups of up to 65x for larger 
explored subgraphs. Even in Peregrine's best case (clique application, 
which contains only one pattern), DuMato is able to deliver consistent speedups. We achieve more expressive gains in motif counting application for larger values of $k$, which can be explained by the inherent characteristics of pattern-aware enumeration of Peregrine. As we increase the length of explored subgraphs, the number of valid patterns and exploration plans grows exponentially, incurring in two aspects that impact the Peregrine performance: (i)~the cost of generating exploration plans for each pattern increases and (ii)~part of exploration plans does not generate valid subgraphs, leading to wasted computational resources.

\section{Conclusion} \label{section:conclusion}

We propose the DuMato GPU based GPM system that integrates novel strategies to address the challenges found in GPM execution on GPUs: \textit{high memory demands}, \textit{memory uncoalescence}, \textit{divergences}, and \textit{load imbalance}. These strategies
and optimizations include our DFS-wide subgraph exploration, warp-centric system design and implementation, and warp-level load balancing strategy. Our system
and optimizations were evaluated using real-world datasets. We compared DuMato to three state-of-the-art GPM systems, showing that it is more scalable (up to 12-vertice subgraphs) and often one order of magnitude faster.

As a future work, we plan to extend our load balancing with a fine-grained asynchronous workload redistribution, allowing work redistribution without having to
stop and restart the GPU kernel. Even with our efficient GPU based design,
the GPM algorithms have a long execution time as we increase the 
length of datasets and explored subgraphs. Thus, we intend to propose a multi-GPU version of DuMato to accelerate it further along with a fault tolerance layer to avoid restarting long runs from scratch.

\bibliographystyle{unsrt}
\bibliography{references}

\end{document}